# Micropillar compression of single crystal tungsten carbide, Part 2: Lattice rotation axis to identify deformation slip mechanisms

Vivian Tong[*], Helen Jones, Ken Mingard

National Physical Laboratory, Hampton Road, Teddington, Middlesex TW11 0LW, UK

## Abstract

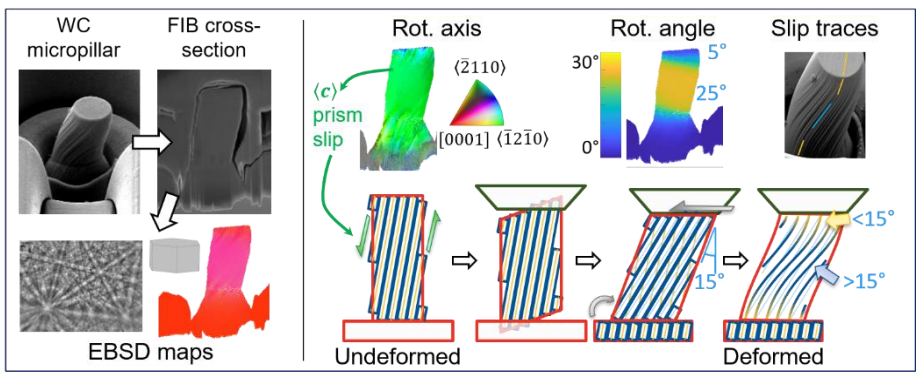

The plastic deformation mechanisms of tungsten carbide at room and elevated temperatures influence the wear and fracture properties of WC-Co hardmetal composite materials. The relationship between residual defect structures, including glissile and sessile dislocations and stacking faults, and the slip deformation activity, which produce slip traces, is not clear. Part 1 of this study showed that $\{10\bar{1}0\}$ was the primary slip plane at all measured temperatures and orientations, but secondary slip on the basal plane was activated at 600 °C, which suggests that $\langle a \rangle$ dislocations can cross-slip onto the basal plane at 600 °C. In the present work, Part 2, lattice rotation axis analysis of deformed WC micropillar mid-sections was used to discriminate $\langle a \rangle$ prismatic slip from multiple $\langle c + a \rangle$ prismatic slip in WC, which enabled the dislocation types contributing to plastic slip to be distinguished, independently of TEM residual defect analysis. Prismatic-oriented micropillars deformed primarily by multiple $\langle c + a \rangle$ prismatic slip at room temperature, but by $\langle a \rangle$ prismatic slip at 600 °C. Deformation in the near-basal oriented pillar at 600 °C can be modelled as prismatic slip along $\langle c \rangle$ constrained by the indenter face and pillar base. Secondary $\langle a \rangle$ basal slip, which was observed near the top of the pillar, was activated to maintain deformation compatibility with the indenter face. The lattice rotations, buckled pillar shape, mechanical data, and slip traces observed in the pillar are all consistent with this model.

## Highlights

- Cross-section EBSD on WC micropillars complements slip trace analysis
- Lattice rotation axis used to find dominant slip direction and plane

---

[*] Corresponding author: vivian.tong@npl.co.uk

- $\langle c+a \rangle$ slip on $\{10\bar{1}0\}$ planes active at room temperature in WC
- $\langle a \rangle$ slip on $\{10\bar{1}0\}$ and $(0001)$ planes active at 600 °C in WC
- Von-Mises criterion fulfilled at 600 °C but not room temperature

# 1 Introduction

Slip trace analysis was used to identify slip planes in single crystal tungsten carbide micropillars compressed at room and high temperatures in Part 1 of this study [1]. This was combined with mechanical stress-strain data to provide insight into deformation and cracking mechanisms of hardmetals in their typical operating conditions as cutting tools. $\{10\bar{1}0\}$ prismatic slip planes were identified from the micropillar slip traces, but the activated dislocation slip systems could not be measured from the slip steps, because slip step vectors on the prismatic plane cannot be unambiguously separated into glide of $\langle a \rangle$, $\langle c \rangle$ and $\langle c+a \rangle$ dislocations [2–5]. In fact, multiple $\langle c+a \rangle$ prismatic slip can produce macroscopic slip steps along $\langle a \rangle$, $\langle c \rangle$, or any direction in between.

The question of whether or not (multiple) $\langle c+a \rangle$ prismatic slip is the only active dislocation slip system, or if $\langle a \rangle$ or $\langle c \rangle$ dislocations slip as well, remains (to the authors' knowledge) still unanswered. This affects our understanding of toughness in WC and WC hardmetal composites: if only $\langle c+a \rangle$ prismatic slip systems are active, only four out of five degrees of freedom required for arbitrary plastic deformation are available, and the toughness of hardmetals comes only from microcracking and binder deformation [6–8]. However, if $\langle a \rangle$ or $\langle c \rangle$ dislocations can cross-slip onto non-prismatic planes, this would enable arbitrary plastic deformation. In Part 1 of our study [1], we observed basal plane slip traces in micropillars compressed at 600 °C, which suggests that $\langle a \rangle$ dislocations can cross-slip onto the basal plane at elevated temperatures.

Transmission electron microscopy (TEM) has been used to characterise residual dislocations in deformed tungsten carbide [2,4,9–13]. Residual dislocation populations can be measured by lattice curvature and used to infer the active slip systems, as glissile dislocations are expected to multiply during deformation (e.g. by Frank-Read sources) [4]. However, inferring plasticity from residual dislocations in WC is not straightforward, because dislocations can also be produced by defect interactions unrelated to plastic deformation. For example, the most common residual defects observed in TEM are glissile $\frac{1}{6}\langle 11\bar{2}3 \rangle$ partial dislocations [4,14], but these can react to form perfect $\langle a \rangle$ and $\langle c \rangle$ dislocations, even in the absence of $\langle a \rangle$ or $\langle c \rangle$ slip. The same partial dislocations can also react to form sessile defects in the deformed structure, even though they did not contribute to plastic deformation [10,12].

In this study, we use electron backscatter diffraction (EBSD)-based grain reference orientation deviation (GROD) mapping to measure lattice rotations in micropillar mid-sections in response to the applied deformation. In a constrained single slip geometry, the GROD rotation axis with respect to the undeformed crystal depends on the activated dislocation type and slip plane. This enables us to characterise the active primary slip system in deformed WC micropillars, independently of residual dislocation analysis methods. In a micropillar geometry, suitable reference orientations can be selected from undeformed regions next to the deformed micropillars.

GROD can also be used to calculate intragranular misorientations within an EBSD map by selecting a reference orientation from the deformed crystal orientations. Intragranular misorientation GROD angles are a useful metric for plastic strain evolution in certain material microstructures and loading geometries [15,16]. However, they are not generally suitable for measuring the crystal rotations due to plastic deformation, because the undeformed orientation cannot be accurately inferred from the deformed grain orientations [17,18]. Intragranular misorientation axis analysis is therefore restricted to characterising grain fragmentation and texture evolution during severe plastic defromation [17], or uncommon deformation modes such as kink banding in hexagonal close-packed alloys [19–22], where large (≥ 6°) lattice rotations are localised to a small part of the grain, and the 'undeformed' crystal orientation can be accessed from the rest of the grain.

The present study uses GROD analysis with respect to the undeformed crystal to study lattice rotations due to plastic deformation. Relatively high angular resolutions of around 0.2° were achieved using 'Refined accuracy' indexing [23]. This enabled the rotation axis to be measured even from small GROD angles, which is not possible with conventional EBSD indexing [24,25]. Details of the EBSD angular resolutions in this work is discussed in Section 7.2.

# 2 Method

## 2.1 Sample preparation

Cylindrical micropillars with a 5 µm top diameter were fabricated using focused ion beam (FIB) in two WC single crystal samples, and compressed at five temperatures between room temperature and 600 °C. One sample was oriented with surface normal near $[10\bar{1}0]$ ('prismatic sample'), and the other sample with surface normal near $[0001]$ ('basal sample'). Slip trace analysis of the deformed pillars was performed to identify the slip planes activated at each deformation temperature. Results of the micropillar deformation and slip trace analysis are described in Part 1 of this study [1].

Three deformed pillars and one undeformed reference pillar were selected for further analysis in the present work; the deformation conditions, measured slip plane traces, and likely slip directions activated in these pillars are summarised in Table 1. Slip systems with high Schmid factor are geometrically better aligned for slip, but the preferred slip systems also depend on the critical resolved shear stresses of different slip systems, which in turn depend on the Peierls energies of the active dislocations on their respective slip planes [26,27].

| Pillar orientation / temperature | Compressive engineering strain | Primary slip plane(s) | Slip direction (Schmid factor) | Secondary slip plane(s) | Slip direction (Schmid factor) |
|---|---|---|---|---|---|
| Prism / RT | 10.5 % | $(\bar{1}100)$ $(\bar{1}010)$ | $[11\bar{2}0]$ (0.42) $[1\bar{2}10]$ (0.38) | | |
| Prism / 600 °C | 5.5 % | $(\bar{1}100)$ $(\bar{1}010)$ | $[11\bar{2}0]$ (0.42) $[1\bar{2}10]$ (0.38) | $(01\bar{1}0)$ | $[0001]$ (0.25) |
| Basal / 600 °C | 8.2 % | $(01\bar{1}0)$ | $[0001]$ (0.21) | $(\bar{1}010)$ $(0001)$ $(\bar{1}100)$ | $[0001]$ (0.17) $[11\bar{2}0]$ (0.22) $[\bar{1}\bar{1}23]$ (0.04) |

*Table 1: Details of the three deformed pillars analysed in this study: engineering compressive strains, activated slip planes measured using slip trace analysis, and the theoretical slip directions determined by maximum Schmid factor.*

Cross-sections of the four pillars were extracted from the bulk sample and mounted on TEM half-grids using an *in situ* FIB lift-out technique. Initial trenching and final polishing steps were performed on a Zeiss Auriga-60 FIB-SEM. The *in situ* lift-out step was performed using an Omniprobe micromanipulator in a FEI/ThermoFisher Helios FIB-SEM.

Figure 1 shows the sample preparation steps. To protect the pillar surface from Ga+ ion damage, a protective SiO2 or Pt coating was deposited on the top and sides of the pillar and over the trench around the pillar, using ion-beam assisted chemical vapour deposition. A cross-section of the entire pillar was lifted out and attached to a copper grid.

One face on each section was polished using 30 kV Ga+ ions at +1.5° glancing incidence, then final polishing using 5 kV Ga+ ions and a +6° incidence angle. The final thicknesses of all cross-sectioned samples were around 1-2 µm.

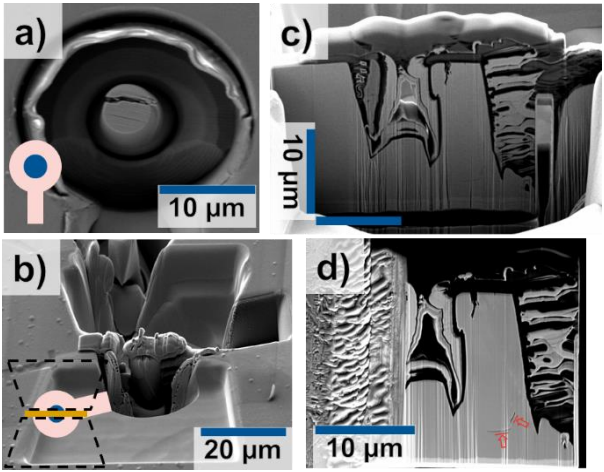

*Figure 1: Pillar cross-section preparation method. a) As-deformed micropillar; b) protective coating deposited and trenches cut around pillar; c) pillar mid-section before FIB lift-out; d) pillar section on a TEM half-grid. Red arrows highlight internal cracks at the pillar base, which were etched by the ion beam during final polishing.*

## 2.2 EBSD

EBSD maps were acquired for all four pillars on a Zeiss Auriga SEM and an Oxford Instruments HKL Nordlys F detector with Aztec software, using 20 kV electron accelerating voltage, and 120 µm aperture size with 'High Current' mode switched on, corresponding to a probe current of ~10 nA. Figure 2 shows a typical EBSD pattern acquired at full detector resolution.

Acquisition parameters for each map are given in Table 2. The field of view for each map contained the entire deformed pillar and some undeformed material at the bottom and sides of the trench. The Aztec software's 'Refined Accuracy' algorithm was used to index crystal orientations. This method refines the conventional Hough-based orientation solution [28] through iterative fitting to Kikuchi band edge positions in the acquired pattern [23]. The mean angular deviation (MAD) is higher, i.e. the orientation precision is lower, in 6×6 than 4×4 binned patterns. The pattern binning levels were chosen purely based on the available microscopy time for these experiments, and were not related to the sample surface finish.

To minimise thermal drift and carbon contamination artefacts, the SEM chamber was plasma cleaned for at least 1 hour before each map, and the electron beam was left to repeatedly scan near an edge of the sample for about 1 hour, until the thermal drift from electron beam heating of the sample had stabilised.

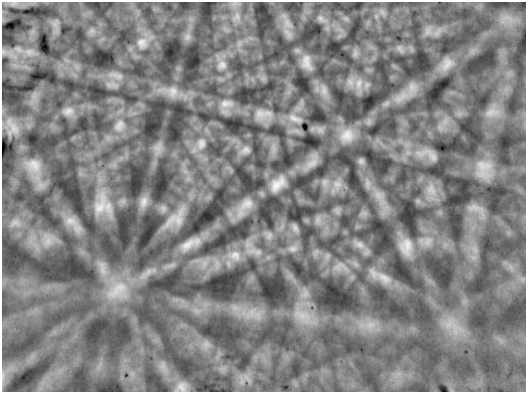

*Figure 2: Full-resolution EBSD pattern acquired from the sectioned micropillars.*

| Sample | Prism undeformed | Basal 600 °C | Prism RT | Prism 600 °C |
|---|---|---|---|---|
| **Working distance** | 13.8 mm | 12.6 mm | 12.9 mm | 12.2 mm |
| **Exposure time** | 28.8 ms | 12.0 ms | 25.0 ms | 19.0 ms |
| **Pattern binning** | 4 × 4 | 6 × 6 | 4 × 4 | 6 × 6 |
| **Step size** | 100 nm | 50 nm | 50 nm | 50 nm |
| **MAD (mode)** | 0.20° | 0.34° | 0.22° | 0.40° |
| **MAD (95th percentile)** | 0.49° | 0.93° | 0.48° | 0.68° |

*Table 2: EBSD map acquisition parameters for the four pillars.*

## 2.3 Data analysis

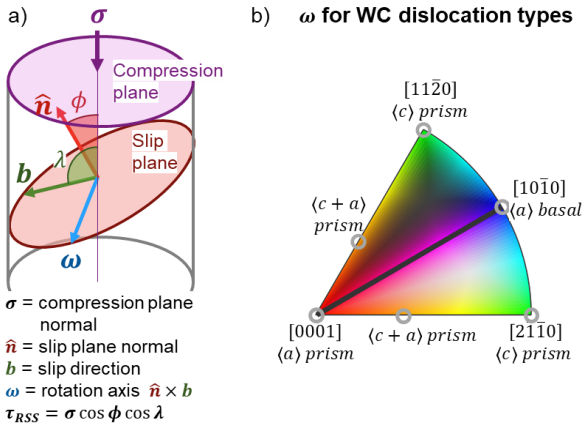

*Figure 3: a) Schematic diagram showing the slip direction, plane, and rotation axis of a sample in compression. $\tau_{RSS}$ is the component of shear stress resolved onto the active slip plane and direction. b) Stereographic projections of theoretical lattice rotation axes for WC dislocation slip on prismatic and basal planes.*

Grain orientation reference deviation (GROD) axis and angle analysis was used to characterise the activated micropillar slip systems. Figure 3a) shows the slip plane, slip direction and rotation axis during compressive deformation. When expressed as crystal coordinates, the GROD axis is characteristic of the active crystallographic slip plane and direction; as sample coordinates, it can be used to characterise slip anisotropy with respect to the pillar compression plane. The rotation angle is correlated to the degree of slip and constraint in the sample [29–31].

The expected rotation axes (in crystal coordinates) for WC single slip systems containing the observed slip planes ($\{10\bar{1}0\}$ and $(0001)$) are plotted on the stereographic projection inverse pole figure (IPF) colour key in Figure 3b). The GROD axes shown are valid for single slip and some special cases of multiple slip in WC.

More information about GROD axis analysis theory, its applications to WC slip, and details of how it has been calculated and plotted in this work, are provided in the Supplementary Information (Section 1).

# 3   Results
## 3.1   Orientations

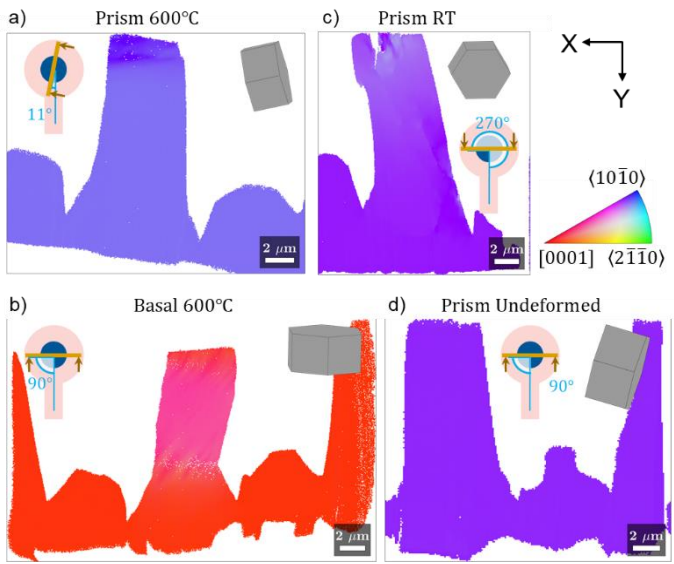

*Figure 4: Orientation maps of the 4 micropillar cross-sections showing inverse pole figure directions along Y, which is approximately parallel to the pillar loading axes. Unit cells show the mean orientation of undeformed regions, which are used as GROD reference orientations.*

Orientation maps of the micropillar cross-sections in Figure 4 show the crystal direction along Y, the approximate pillar loading direction.

Orientation gradients are visible as colour changes in the three deformed pillars. Hexagonal unit cells in the top right corner of each map show the undeformed crystal orientations used as GROD references. Differences in unit cell orientations of the three prismatic pillars are due their different sectioning planes, as shown by keyhole-shaped schematics in Figure 4.

## 3.2 GROD angles

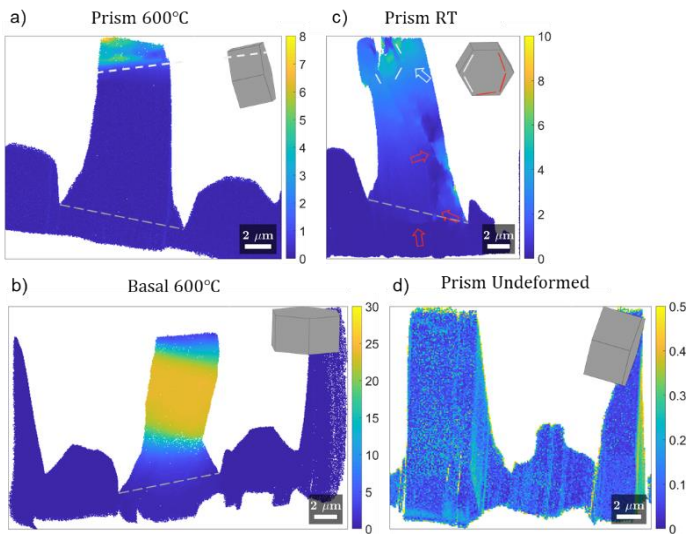

*Figure 5: Grain reference orientation deviation (GROD) angle maps. Colour scale units are in degrees; the large change in scale range between the different maps should be noted. The bottom of the pillar is defined as where the pillar walls meet the base, marked by dashed grey lines in subfigures a)-c).*

GROD rotation angles are shown in Figure 5. High rotation angles correspond to high degrees of constrained slip.

Vertical line features < 0.3° are visible in the undeformed pillar (Figure 5d); these are curtaining topography artefacts from FIB sample preparation.

Figure 5a shows the prismatic pillar deformed at 600 °C. The deformation has produced a misoriented region of 4 - 8°, localised to the top 1-2 µm of the pillar. The misoriented region is marked by a diffuse boundary with surface trace near [0001] (white dashed lines), below which lattice rotations are negligible.

Figure 5b shows the basal pillar deformed at 600 °C. The middle section of the pillar has uniformly rotated about 25° from the undeformed crystal, but the top and bottom of the pillar have rotated only about 5°. The misorientation angle increases smoothly from 5° to 25° in the top and bottom 2 µm of the pillar. The deformed pillar is tilted towards the right.

Figure 5c shows the prismatic pillar deformed at room temperature. Long-range lattice rotations between 1° and 4° are visible along the left-hand side of the pillar, where the rotation angle decreases smoothly from top to bottom.

A large crack is present at the top left of the pillar, where a corner of the pillar sheared significantly along a $\langle 11\bar{2}0 \rangle$ direction and in a $\{1\bar{1}00\}$ plane. Equiaxed subgrains (white arrow) with GROD angle around 6° are visible on the right side of the large crack near the top of the pillar. Some subgrain boundary lines are faceted along $\langle 11\bar{2}0 \rangle$ surface traces (white lines), and might be $\{1\bar{1}00\}$ plane traces.
Three orientation discontinuities, with surface traces near $\langle 11\bar{2}0 \rangle$, are visible on the lower-right side of the pillar and in line with the pillar base (red arrows on map, red lines in unit cell). The discontinuities near the pillar base are two cracks revealed by FIB etching during sample preparation (red lines and arrows in Figure 1d). The material directly underneath the crack along the base shows negligible lattice rotation.

## 3.3 Misorientation axes in crystal coordinates

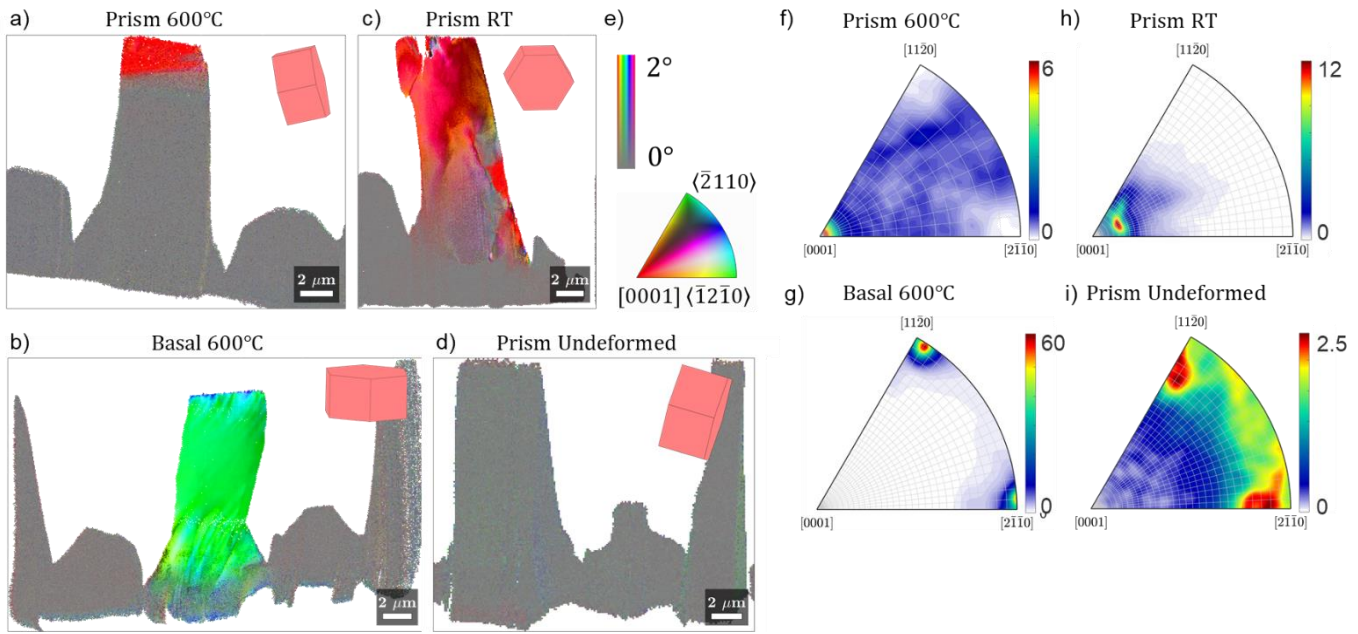

*Figure 6: GROD axes in crystal coordinates. a-d) GROD axis maps of the four samples coloured according to the spherical axis-angle colour key in e). f-i) Distribution of GROD axis vectors in the pillar regions, colour scaling shows frequency in multiples of a uniform spherical distribution; stereographic projection grid spacing is 6° in f) and 3° in g-i).*

GROD axes in crystal coordinates, plotted in Figure 6, are related to the characteristic lattice rotation axes of the activated slip systems (Figure 3). The GROD axis colour key is based on Reference [32] and implemented in MTEX. This plots the misorientation axis in crystal coordinates using a direction colour key based on the stereographic projection IPF triangle [33].

At larger rotation angles > 2°, the rotation axis is well defined, and map colours are equivalent to the IPF colours. For small misorientations < 2°, the rotation axis is uncertain, so the colour saturation decreases to reflect this. Saturation decreases linearly between 2° and 0° misorientation, which is grey for any misorientation axis, consistent with the fact that the misorientation axis is fully degenerate when there is no misorientation.

Figure 6 provides initial confirmation that 2° is a suitable threshold value, since deformation-related structures in the pillars have high colour saturation, whilst undeformed regions have low saturation and appear grey. Formal justification for choosing a 2° misorientation threshold is provided in the Appendix (Section 7.2). IPF direction maps of the GROD axis only, without considering degeneracy at small angles, are shown in Supplementary Figure 2 (crystal coordinates) and Supplementary Figure 3 (sample coordinates).

The undeformed pillar map (Figure 6d) is uniformly grey, because the measured misorientations are only from noise and topography artefacts. These artefacts appear as peaks near $\langle 11\bar{2}0 \rangle$ in the pole figure (Figure 6i).

The GROD axis of the 600 °C deformed prismatic pillar (Figure 6a,f) points along [0001], corresponding to $\langle a \rangle$ prismatic slip. Only the top 1-2 μm of the pillar has rotated. The pole figure peak (Figure 6f) is narrow (< 6°) and centred on [0001], indicating that only $\langle a \rangle$ prismatic slip was activated on two prismatic planes (identified by slip trace analysis in Part 1 of this study [1]). The diffuse boundary between red and grey runs along [0001], and is most likely a $\{10\bar{1}0\}$ slip plane trace.

The GROD axis of the room temperature deformed prismatic pillar (Figure 6c, h) points near [0001]. However, the EBSD maps show orange and pink IPF colours, and the pole figure (Figure 6h) shows that the measured GROD axis is oriented about 15 ° away from [0001] towards ⟨10$\bar{1}$0⟩.

The 600 °C deformed basal pillar (Figure 6b) has a GROD axis of ⟨11$\bar{2}$0⟩ in most of the pillar, corresponding to prismatic slip with total slip direction along ⟨c⟩. Slip trace analysis of this sample [1] showed that slip primarily occurred on prismatic planes, and basal slip was a minor slip system near the top of the pillar. In some regions near the top of the pillar, the misorientation axis changes to ⟨10$\bar{1}$0⟩ (blue). This change might have been caused by local secondary slip activity, as the stress distribution near the top of the micropillar is extremely sensitive to indenter alignment, pillar edge rounding, and other geometric parameters that vary significantly from pillar to pillar [34].

### 3.4 Misorientation axes in sample coordinates

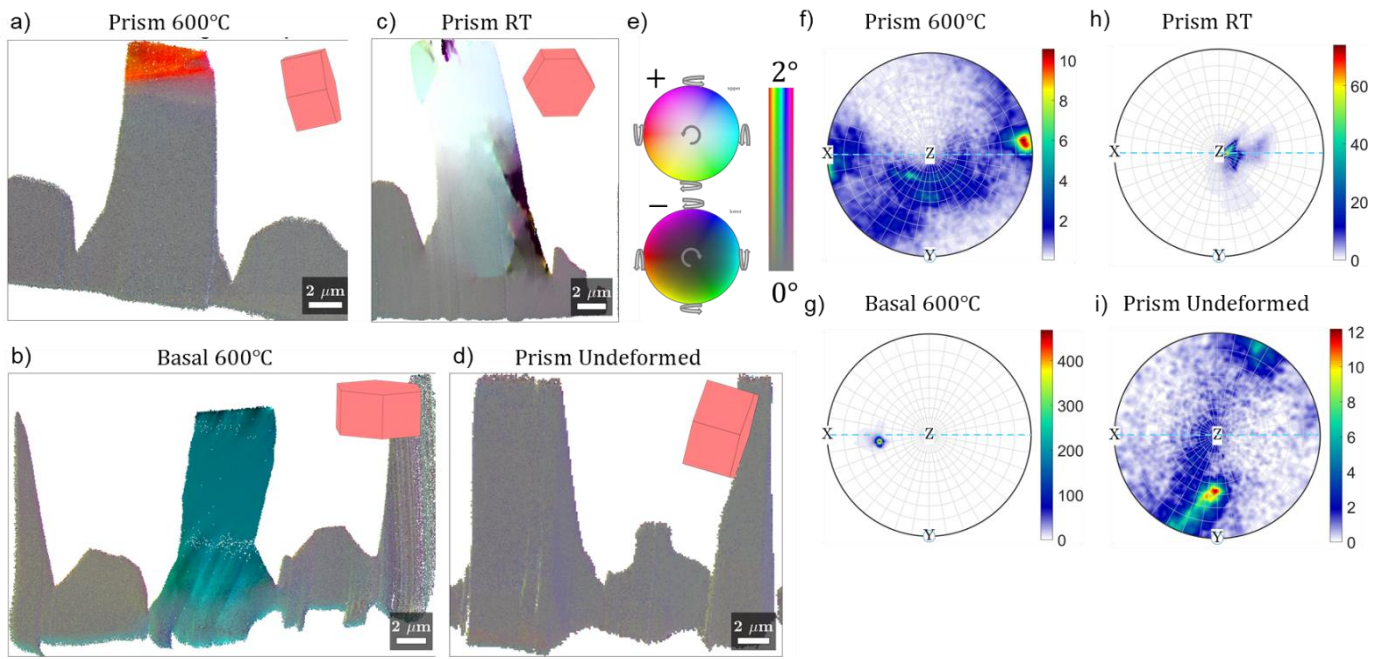

*Figure 7: GROD axes in sample coordinates. Consistent coordinate axes are used to plot the EBSD maps (a-d) and corresponding pole figures of the northern hemisphere (f-i). The GROD colour keys for sample coordinates is shown in (e) with a threshold rotation angle of 2°. The stereographic projection grid line spacing in the pole figures is 10°.*

GROD axes in sample coordinates are mapped in Figure 7a-d using a spherical colour key which includes both hemispheres to distinguish opposite signed rotations, and is based on a HSV (Hue-Saturation-Value) colourspace. Colour saturation decreases for points with rotation angle < 2°, similar to Figure 6a-d. A right-handed rotation convention is used, i.e. anticlockwise when the rotation axis points out of the page. A white rotation axis points out of the page, representing an anticlockwise in-plane rotation. A black rotation axis points into the page and represents a clockwise in-plane rotation. Colour hue changes with the out-of-plane rotation component, as the equator cycles around 'rainbow' colours. The circular arrow annotations in Figure 7e illustrate rotation directions around the equator and at the poles. Figure 7f-i shows pole figures of the same data, but with all points projected into the northern hemisphere (i.e. points on the southern hemisphere are inverted).

All deformed pillars had rotation axes nearly perpendicular to the loading direction. The plane perpendicular to the loading direction or pillar axis is marked in the pole figures (Figure 7f-i) by blue dashed lines. Approximately 5° uncertainty in loading direction can be expected from sample misalignment and taper from FIB cross-sectioning and lift-out; a detailed justification is given in the Appendix.

In the 600 °C deformed prismatic pillar, Figure 7a shows that out-of-plane lattice rotations are present only at the top 1-2 µm of the pillar, with the top of the pillar rotating into the page about the +X-axis.

In the 600 °C deformed basal pillar, Figure 7b shows a uniform lattice rotation axis in the middle of the pillar, which points 80° from the loading direction and 40° out of the sectioning plane.

In the room temperature deformed prismatic pillar, Figure 7c shows that the lattice rotation axes were nearly perpendicular to the pillar cross-sectioning plane. On either side of the crack and slip lines on the pillar, the crystal rotated to widen or open the crack, which is shown by the anticlockwise rotation (white-green) on the left side of the slip line or crack, and clockwise (black-purple) on the right.

In the undeformed pillar, Figure 7i shows a narrow peak in the pole figure, which is a topography artefact due to a slight sample bevel during final FIB polishing. The sample bevel introduced a systematic change in the EBSD pattern brightness, producing an indexing artefact which is hardly visible in the EBSD map in Figure 7d, but can be seen in Supplementary Figure 3b as a blue stripe down the right side of the pillar.

# 4 Discussion

The primary slip system determined from the GROD axis for all three deformed pillar cross-sections agrees with slip trace analysis in part 1 of this study [1].

## 4.1 Prismatic pillar, RT deformation

### 4.1.1 Slip

Both slip trace analysis [1] and microstructural features in the GROD axis maps (Figure 6 and Figure 7) show that two prismatic slip systems with total slip direction near $\langle a \rangle$ were activated in the pillar.

In this section, detailed analysis of the GROD rotation axis will be used to determine whether prismatic slip in this pillar was produced by $\langle a \rangle$ or multiple $\langle c + a \rangle$ dislocation types. $\langle a \rangle$ dislocations limit slip to $\langle a \rangle$ directions and produce a GROD axis close to $[0001]$. In contrast, the total slip direction for multiple $\langle c + a \rangle$ dislocation slip is not restricted to any particular direction in the prismatic plane, but can be determined instead by the pillar loading stress. The most favourable slip direction, which is along the maximum resolved shear stress direction, results in a GROD axis perpendicular to the loading direction, and lying in the prismatic plane.

In the two deformed prismatic pillars, the angles between the maximum resolved shear stress directions and nearest $\langle a \rangle$ directions were 17° for both of the activated prismatic slip planes. Therefore, the GROD axis for multiple $\langle c + a \rangle$ slip is expected to deviate from $[0001]$ by approximately 17°.

Figure 6 and Figure 7 show that the RT prismatic pillar deformed by multiple $\langle c + a \rangle$ prismatic slip, not $\langle a \rangle$ slip: the GROD axis in crystal coordinates (Figure 6h) deviates from $[0001]$ by about 15° towards $\langle 10\bar{1}0 \rangle$, and the GROD axis in sample coordinates (Figure 7h) is nearly perpendicular to the loading direction Y. These results are significant within their uncertainty bounds (details are in the Appendix): 3-7° sample orientation uncertainty and ≤ 6° GROD axis uncertainty.

Therefore, even though the total slip direction was near $\langle a \rangle$, it was likely produced by cooperative multiple $\langle c + a \rangle$ dislocation slip. Deviation of the GROD axis peak from the red-orange-green line between $[0001]$ and $\langle 11\bar{2}0 \rangle$ indicates that $\langle c + a \rangle$ dislocation slip was activated on multiple prismatic planes.

### 4.1.2 Cracking

A wide crack on the top left of the pillar has grown subsurface along a {10-10} plane, close to slip bands with severe shear localisation. This means the crack plane most likely opened after slip was exhausted, i.e. no more glissile dislocations were available on this {10-10} plane, all of them having been pinned by other defects or emitted at the pillar surface. This is consistent with literature observations of high defect density on {10-10} planes, including sessile defects such as stair-rod dislocations from partial dislocation reactions [10], which can also pin other dislocations.

Grain fragmentation is visible in the GROD maps, with oppositely signed rotations on either side of the subgrain boundaries (Figure 7c) which could have opened cracks in the pillar on further loading. The subgrain boundaries are sections through {10-10} slip traces on the pillar surface, and two cracks can be observed near the base of the pillar with a shadow on the bottom face from ion-beam etching (red arrows in Figure 1d). The crack traces lie along $\langle 1\bar{2}10 \rangle$ and the crack planes are likely {10-10} planes. The cracks mark a discontinuity in the GROD maps: in the near-horizontal crack at the pillar base, the GROD angle above the crack line is 0.5°, but negligible (< 0.2°) below the crack.

## 4.2 Prismatic pillar, 600 °C deformation

The primary slip system determined by GROD rotation axis analysis was $\langle a \rangle$ prismatic slip. SEM slip trace analysis showed double slip on two prismatic slip planes, with slip bands intersecting at the top 3-4 μm of the sample.

The active slip planes in this sample are the same as the RT deformed prismatic pillar (Section 4.1). However, this pillar was cross-sectioned almost parallel to the intersection line between the two slip planes (Figure 4a), so that lattice rotation contributions from the two slip systems could not be observed separately, as they were for the RT deformed pillar.

This pillar is the same orientation as the prismatic pillar deformed at RT, so the same analysis in Section 4.1.1 has been applied to distinguish between $\langle a \rangle$ slip versus cooperative $\langle c + a \rangle$ slip. At 600 °C, the pillar deformed by $\langle a \rangle$ dislocation slip on prismatic planes. The GROD axis peak is within 1° of [0001] in crystal coordinates (Figure 6f), but points out of the loading plane by about 8° in sample coordinates (Figure 7f), i.e. the chosen slip direction was not the maximum shear stress direction.

SEM imaging (Figure 6 of [1]) showed cracks on the pillar top face propagating down into the pillar. The crack paths jump between both sets of $\{1\bar{1}00\}$ slip bands on the pillar top face, and extends down the pillar length between the two slip planes. This suggests that the two $\langle a \rangle$ prismatic slip systems operated at the same time, not one after the other.

Only the top 1-2 μm of material has rotated relative to the undeformed orientation in this pillar cross-section. This is consistent with the pillar slip traces, which show that slip is localised to the top part of the pillar near the pillar mid-section, where the two sets of slip bands intersect. The lattice rotation axes in sample coordinates shows the pillar rotating backwards into the page with a horizontal rotation axis, corresponding to the primary slip system operating in this sectioning plane. The observed 'primary' slip system likely switches through the pillar depth. If the FIB cross-sectioning plane had been slightly shallower or deeper through the pillar, the other double $\langle a \rangle$ prismatic slip system could have been observed as the primary slip system.

## 4.3 Basal pillar, 600 °C deformation

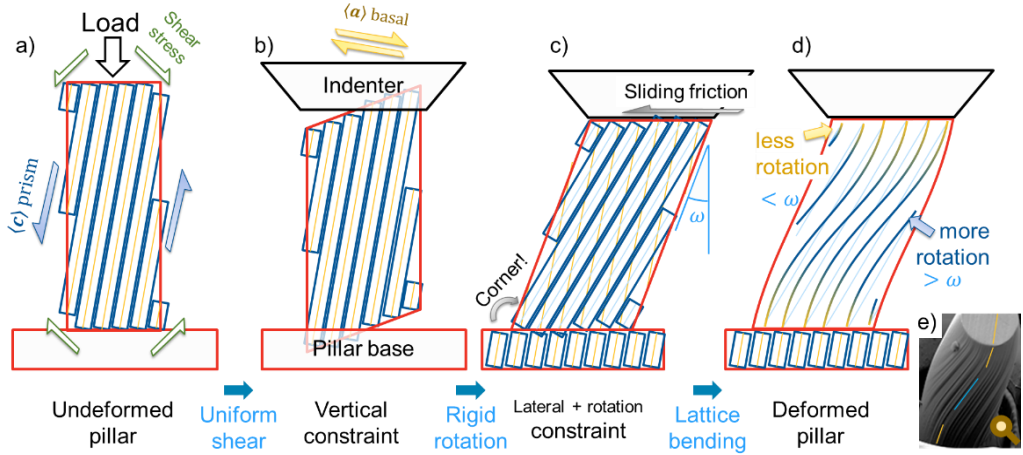

Figure 8: Schematic of ⟨c⟩ prismatic slip deformation in the basal pillar at 600°. The schematic is drawn so that primary slip plane runs perpendicular in and out of the the page. The slip direction ⟨c⟩ lies in the plane of the page. The angle between the prismatic slip plane normal and the pillar loading direction is around 80°; this is drawn to scale in the figure.

The primary slip system previously assumed from slip trace and Schmid factor analyses was ⟨c⟩ prismatic slip. For this crystal orientation, the GROD axis could not be used to distinguish ⟨c⟩ dislocation slip from cooperative multiple slip of ⟨c + a⟩ dislocations, because the maximum resolved shear stress direction on the prismatic plane deviates only 5° from ⟨c⟩. The GROD axis expected from these two deformation modes would be indistinguishable within the experimental uncertainty.

The high lattice rotation angles indicate that slip was constrained during deformation, but the tilted shape of the deformed pillar curved slip traces show significant plastic rotations as well, which is expected in an unconstrained slip deformation geometry. Therefore, an approximate model, described in Section 1.3 of the Supplementary Information, and similar to the micropillar plastic buckling schematic in Reference [35], was used to check if the lattice rotations measured by EBSD, compressive strains from the mechanical data, and the deformed pillar shape observed by SEM imaging, were consistent with each other.

Figure 8 shows how a uniaxial compressive strain can be decomposed into ⟨c⟩ prismatic shear and rotation components for this pillar's crystal orientation and loading geometry. Figure 9 then quantifies the theoretical shear and rotations expected from the 8 % compressive strain (Table 1) applied to this pillar.

Figure 8a shows a schematic diagram of the primary slip system with respect to the loading axis. Both the primary slip plane and slip direction are oriented almost vertically in the pillar, therefore shear on this plane is vertically constrained by the pillar base and flat punch indenter (Figure 8b), which has been accommodated by a combination of plastic deformation and rigid body rotation, as no cracking was observed in this sample.

⟨a⟩ basal slip is theoretically well-oriented to accomodate the indenter constraint (Figure 8b), and experimentally confirmed by basal slip traces observed near the top of the pillar (Table 1) in Part 1 of the study . The active dislocation type is most likely ⟨a⟩ screw, although this was not experimentally measured. We inferred this from several facts: that ⟨c + a⟩ dislocations cannot slip on the basal plane, but ⟨a⟩ screw dislocations can cross-slip onto the basal plane, and that ⟨a⟩ dislocations are glissile in WC at 600 °C, as they were observed in the prismatic pillar (Section 4.2).

The vertical constraint can also be accommodated by rigid body rotation of the entire pillar by the angle ω, as shown in Figure 8c. However, the pillar rotation is partially laterally constrained by indenter friction, to prevent the pillar top face from sliding. It is laterally and rotationally constrained at the pillar base, where the pillar is continuous with the base material, and a discontinuity in the crystal lattice would have led to cracking, which was not observed here. Figure 8d shows the lattice rotation gradients required to accommodate the geometric constraints in Figure 8c at the top and base of the pillar.

The experimentally observed shape of the deformed pillar qualitatively matches the schematic drawing in Figure 8d. Figure 8e shows the deformed pillar imaged from an equivalent viewing orientation. The yellow and blue lines mark the slip trace orientations on the top face and side of the pillar.

Another way of maintaining vertical compatibility with the flat indenter tip in Figure 8b, besides a rigid body rotation, is by activating secondary slip systems such as $\langle a \rangle$ basal slip. This is consistent with the basal slip traces observed in Part 1 of this study. However, this is a tentative suggestion, as the local stress at the top of the pillar is extremely sensitive to pillar shape and indenter alignment [34], and these contributions were not quantified for this particular pillar.

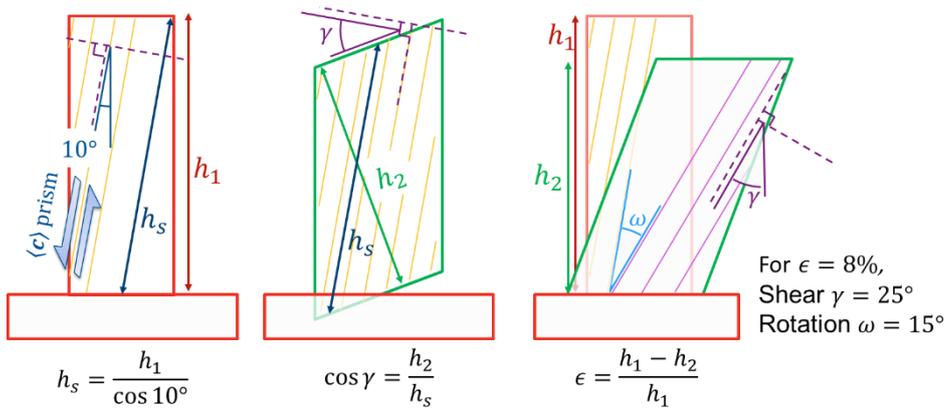

Figure 9: Schematic showing the approximate lattice rotation angle expected in the basal pillar, which was deformed by $\langle c \rangle$ prismatic slip to produce 8% engineering compressive strain ($\epsilon$). The total deformation is decomposed into shear ($\gamma$) and rigid body rotation ($\omega$) components. $h_1$ and $h_2$ are the initial and undeformed pillar heights respectively, and $h_s$ is the pillar height resolved onto the slip direction $\langle c \rangle$.

Figure 9 shows a 2D approximation of the theoretical lattice rotation ω and shear strain γ expected for an engineering compressive strain ϵ, which was measured from the mechanical loading data to be 8 %. Achieving 8% compressive strain through $\langle c \rangle$ prismatic slip requires 25° shear strain (γ) followed by 15° lattice rotation (ω). The qualitative lattice bending shown in Figure 8d then leads us to expect > 15° lattice rotation in the middle of the pillar, and < 15° near the top and base. This is consistent with the experimental data (Figure 5), which shows 25° lattice rotation in the middle of the pillar and 5° near the top and base.

The angle between prismatic slip planes and the loading direction was 10° before deformation, but 35° afterwards. This means that the resolved shear stress on the $\langle c \rangle$ prismatic slip system increased during deformation, so that the pillar would appear to soften as it deformed. This confirms that the first large load drop observed in Part 1 of this study for basal-oriented pillars is due to unstable yield, caused by geometric softening of the $\langle c \rangle$ prismatic slip system [1].

Analogous behaviour has been observed in room temperature nanoindentation by Csanádi et al. [36]. They observed rapid softening as the $\langle c \rangle$-axis declination angle increased from 20° to 40°, then a plateau for angles > 40°, and modelled this anisotropy simply by considering geometrical alignment of $\langle c + a \rangle$ prismatic slip systems. This orientation dependence is consistent with the yield softening observed in the present study. The plateau in indentation hardness beyond 40° is also consistent with the $\langle c \rangle$ declination angle of 35° in the deformed basal pillar, as yield is expected to stabilise at around this point.

# 5 Summary

GROD analysis of deformed micropillars has been used to characterise lattice rotations in deformed WC micropillars, to identify the primary slip systems activated in WC at room and high temperatures. In particular, this method enabled $\langle a \rangle$ dislocation slip to be distinguished from cooperative multiple $\langle c + a \rangle$ dislocation slip on prismatic planes, independently of any residual defect analysis.

We found that cooperative multiple $\langle c + a \rangle$ prismatic slip was the primary room temperature slip system, consistent with published TEM residual dislocation analyses [2,4,9–13]. However, at 600 °C, the primary dislocation slip system switched to $\langle a \rangle$ prismatic slip. The switch in active dislocation type to $\langle a \rangle$ dislocations is consistent with the appearance of basal plane slip traces at 600 °C, which was reported in Part 1 of this study.

Since distinguishing $\langle a \rangle$ from multiple $\langle c + a \rangle$ prismatic slip requires interpretation of relatively small changes (< 20°) in GROD axis, a detailed uncertainty analysis of EBSD orientations and misorientations was performed to assess the physical validity of our analysis. A median indexing precision of 0.2°, achieved using the Oxford Instruments refined indexing algorithm, enabled a median GROD axis uncertainty of 6° for a cut-off rotation angle of 2°.

The large lattice rotation angles and bent shape of the basal oriented pillar deformed at 600 °C were caused by the activated prismatic slip planes being constrained by the indenter face and pillar base, and could be approximately quantified by decomposing the compressive deformation into $\langle c \rangle$ prismatic shear and rigid body rotation components. Slip on (0001) planes was observed as a secondary slip system, and may have been activated to maintain deformation compatibility between the indenter and sample face.

# 6 Acknowledgements

Thank you to Phillip Crout and Simon Griggs, for access to and assistance with the Helios FIB at the University of Cambridge Wolfson Electron Microscopy Suite. Thank you to the Department of BEIS (Department for Business, Energy and Industrial Strategy) for funding, under the NMS (National Measurement System) program. Thank you to Raj Ramachandramoorthy (Empa, now at MPIE Düsseldorf) for carrying out the micropillar compression testing. Thank you to Hannah Zhang (NPL) and Mark Gee (NPL) for helpful comments during internal review of the manuscript before submission.

# 7 Appendix

## 7.1 Sample orientation uncertainty

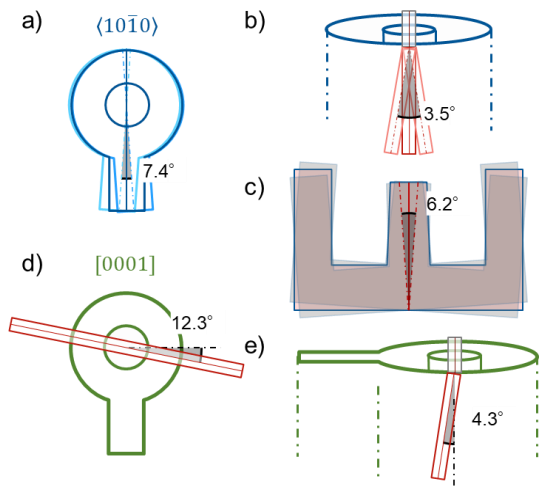

*Figure 10: Scatter in measured sample orientations in the four cross-sectioned micropillars, used to estimate absolute orientation accuracy. (a-c) Range in measured orientations in the three prismatic pillar cross-section and prismatic bulk crystal surface. (d,e) Orientation difference between the basal pillar cross-section and basal bulk crystal surface.*

Uncertainty in cross-sectioning plane due to stage misalignments and to taper during FIB sample preparation was estimated by comparing the orientations between different cross-sectioned samples, and between cross-sectional and surface EBSD measurements [1]. Orientation errors can accumulate from a tilted sample surface in the bulk crystal, stage and sample positioning errors during FIB lift-out, taper and misalignment during FIB cross-sectioning, and in-plane sample rotation of the TEM half-grid in its sample holder.

The measurement scatter of pillar axis directions in the EBSD maps was between 3.5° and 6.2°, shown schematically in Figure 10b, c and e for the two sample orientations. This indicates an estimated orientation uncertainty in pillar loading axis of 3 - 6°. The orientation error around the pillar axis (Figure 10a and d) is up to 12°, but this orientation error component does not affect the pillar loading direction in the EBSD data.

The apparent in-plane rotations of the pillar EBSD maps (e.g. Figure 4a) are a combination of genuine in-plane rotations (Figure 10c), and image skew artefacts from out-of-plane sample tilts [37]. Since it is not trivial to separate these two effects, the EBSD maps were not re-aligned according to the pillar shape or other spatial features.

The orientation errors of up to 12° around the pillar axes in Figure 10a and d most likely resulted from misaligned surfaces in the prismatic and basal-oriented bulk WC crystals respectively. The surface misalignment angle of the bulk WC crystals, and the expected in-plane rotation artefacts, were estimated using the stage rotation method in Reference [37]. The surface misalignment angle was about 2.3° in both samples, leading to expected artefacts of 3° at 52-54° FIB tilt and 6° at 70° EBSD tilt angles. Since the surface EBSD data was acquired at 70° stage tilt, and the pillar cross-sections were fabricated over multiple sessions on two different FIB-SEM instruments, a total in-plane orientation error around the pillar axis is up to 12° seems reasonable. Fortunately, the in-plane orientation error does not affect the validity of GROD axis analysis in the sample frame, as these were interpreted with respect to the loading direction only.

## 7.2 EBSD angular precision

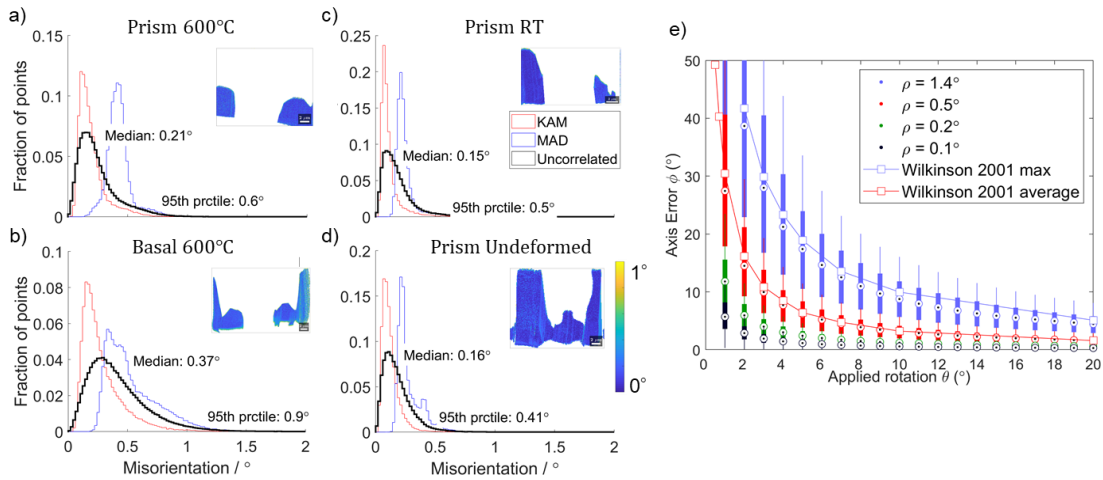

*Figure 11: (a-d) EBSD angular resolution parametrised as KAM (red), MAD (blue) and spatially uncorrelated misorientation (black) distributions, in the undeformed regions of all micropillars. Inset: EBSD map regions considered in each histogram plot, showing GROD angle w.r.t. mean undeformed orientation. e) Box and whisker plots showing theoretical misorientation axis uncertainty as a function of applied rotation angle $\vartheta$, for a range of simulated measurement errors $\rho$ (coloured groups).*

The GROD axis has been used to characterise the crystallographic slip systems which produced lattice rotations. However, EBSD measurement uncertainty can limit interpretation of the GROD axis. Early studies [24,25] showed that the rotation axis uncertainty increases as applied rotation angle decreases. For a 2° applied rotation, the rotation axis error is on average 15° and at worst 40° [25]. However, the EBSD data presented here was indexed by an algorithm which has better angular resolution than conventional Hough-based methods [23]. Therefore, the angular resolution of this data has been considered explicitly.

EBSD data from undeformed regions were used to check the orientation precision. Details of the method are provided in Section 2 of the Supplementary Information file. Figure 11a-d show histogram distributions of three EBSD angular resolution metrics in the undeformed regions of each pillar: mean angular deviation (MAD), kernel average misorientation (KAM), and spatially uncorrelated misorientation distributions. The spatially uncorrelated misorientation distribution (black lines) was chosen to parametrise the orientation uncertainty. Histograms of the three prismatic oriented pillars (Figure 11a, b, d) show a median orientation scatter of about 0.2° and 95$^{th}$ percentile of 0.5°.

Figure 11e plots the theoretical rotation axis error as a function of applied rotation angle, for different amounts of measurement uncertainty. At a 2° applied rotation, which is the threshold angle used for the colour keys in Figure 6 and Figure 7, the expected rotation axis uncertainty is ≤ 6° for 0.2° measurement uncertainty (green box plots) and ≤ 15° for 0.5° measurement uncertainty (red box plots).

This can now be applied to the micropillar EBSD maps, which have an orientation measurement uncertainty of ≤ 0.2° in 50 % of points, and ≤ 0.5° in 95 % of points (Figure 11a, b, d). Therefore, if the GROD angle is 2°, the corresponding uncertainty in GROD axis is ≤ 6° for 50 % of points, and ≤ 15° for 95 % of points. Points with GROD angles > 2° have a smaller GROD axis uncertainty. This level of sensitivity was enough to identify the primary dislocation slip systems of each pillar, including distinguishing $\langle a \rangle$ from multiple $\langle c + a \rangle$ dislocation slip (Section 4.1.1, 4.2).

# 8 References

[1] H. Jones, V. Tong, R. Ramachandramoorthy, K. Mingard, J. Michler, M. Gee, Micropillar compression of single crystal tungsten carbide, Part 1: temperature and orientation dependence of deformation behaviour, n.d.

[2] X. Liu, J. Zhang, C. Hou, H. Wang, X. Song, Z. Nie, Mechanisms of WC plastic deformation in cemented carbide, Mater. Des. 150 (2018) 154–164. https://doi.org/10.1016/j.matdes.2018.04.025.

[3] S.B. Luyckx, Slip system of tungsten carbide crystals at room temperature, Acta Metall. 18 (1970) 233–236. https://doi.org/10.1016/0001-6160(70)90028-3.

[4] M.K. Hibbs, R. Sinclair, Room-temperature deformation mechanisms and the defect structure of tungsten carbide, Acta Metall. 29 (1981) 1645–1654. https://doi.org/10.1016/0001-6160(81)90047-X.

[5] T. Takahashi, E.J. Freise, Determination of the slip systems in single crystals of tungsten monocarbide, Philos. Mag. 12 (1965) 1–8. https://doi.org/10.1080/14786436508224941.

[6] V. Jayaram, Plastic incompatibility and crack nucleation during deformation on four independent slip systems in tungsten carbide-cobalt, Acta Metall. 35 (1987) 1307–1315. https://doi.org/10.1016/0001-6160(87)90012-5.

[7] V. Jayaram, R. Sinclair, D.J. Rowcliffe, Intergranular cracking in WC-6% Co: An application of the von mises criterion, Acta Metall. 31 (1983) 373–378. https://doi.org/10.1016/0001-6160(83)90214-6.

[8] B. Roebuck, E.A. Almond, Deformation and fracture processes and the physical metallurgy of WC-Co hardmetals, Int. Mater. Rev. 33 (1988) 90–112. https://doi.org/10.1179/imr.1988.33.1.90.

[9] J.D. Bolton, M. Redington, Plastic deformation mechanisms in tungsten carbide, J. Mater. Sci. 15 (1980) 3150–3156. https://doi.org/10.1007/BF00550388.

[10] S. Lay, HRTEM investigation of dislocation interactions in WC, Int. J. Refract. Met. Hard Mater. 41 (2013) 416–421. https://doi.org/10.1016/j.ijrmhm.2013.05.017.

[11] R.M. Greenwood, M.H. Loretto, R.E. Smallman, The defect structure of tungsten carbide in deformed tungsten carbide-cobalt composites, Acta Metall. 30 (1982) 1193–1196. https://doi.org/10.1016/0001-6160(82)90013-X.

[12] S. Hagege, J. Vicens, G. Nouet, P. Delavignette, Analysis of structure defects in tungsten carbide, Phys. Status Solidi. 61 (1980) 675–687. https://doi.org/10.1002/pssa.2210610242.

[13] T. Johannesson, B. Lehtinen, The analysis of dislocation structures in tungsten carbide by electron microscopy, Philos. Mag. 24 (1971) 1079–1085. https://doi.org/10.1080/14786437108217070.

[14] M.K. Hibbs, R. Sinclair, D.J. Rowcliffe, Defect Structure of WC Deformed at Room and High Temperatures, in: R.K. Viswanadham, D.J. Rowcliffe, J. Gurland (Eds.), Sci. Hard Mater., Springer US, Boston, MA, 1983: pp. 121–135. https://doi.org/10.1007/978-1-4684-4319-6_7.

[15] A. Harte, M. Atkinson, M. Preuss, J. Quinta da Fonseca, A statistical study of the relationship between plastic strain and lattice misorientation on the surface of a deformed Ni-based superalloy, Acta Mater. 195 (2020) 555–570. https://doi.org/10.1016/j.actamat.2020.05.029.

[16] S.S. Rui, L.S. Niu, H.J. Shi, S. Wei, C.C. Tasan, Diffraction-based misorientation mapping: A continuum mechanics description, J. Mech. Phys. Solids. 133 (2019). https://doi.org/10.1016/j.jmps.2019.103709.

[17] J.Y. Kang, B. Bacroix, H. Réglé, K.H. Oh, H.C. Lee, Effect of deformation mode and grain orientation on misorientation development in a body-centered cubic steel, Acta Mater. 55 (2007) 4935–4946. https://doi.org/10.1016/j.actamat.2007.05.014.

[18] V. Tong, E. Wielewski, B. Britton, Understanding plasticity in zirconium using in-situ measurement of lattice rotations, ArXiv. (2019) 15.

[19] Y. Zheng, W. Zeng, Y. Wang, S. Zhang, Kink deformation in a beta titanium alloy at high strain rate, Mater. Sci. Eng. A. 702 (2017) 218–224. https://doi.org/10.1016/j.msea.2017.07.015.

[20] M. Yamasaki, K. Hagihara, S.I. Inoue, J.P. Hadorn, Y. Kawamura, Crystallographic classification of kink bands in an extruded Mg-Zn-Y alloy using intragranular misorientation axis analysis, Acta Mater. 61 (2013) 2065–2076. https://doi.org/10.1016/j.actamat.2012.12.026.

[21] T. Mayama, T. Ohashi, Y. Tadano, K. Hagihara, Crystal Plasticity Analysis of Development of Intragranular Misorientations due to Kinking in HCP Single Crystals Subjected to Uniaxial Compressive Loading, 56 (2015) 963–972.

[22] T. Matsumoto, M. Yamasaki, K. Hagihara, Y. Kawamura, Configuration of dislocations in low-angle kink boundaries formed in a single crystalline long-period stacking ordered Mg-Zn-Y alloy, Acta Mater. 151 (2018) 112–124. https://doi.org/10.1016/j.actamat.2018.03.034.

[23] K. Thomsen, N.H. Schmidt, A. Bewick, K. Larsen, J. Goulden, Improving the Accuracy of Orientation Measurements using EBSD, Microsc. Microanal. 19 (2013) 724–725. https://doi.org/10.1017/s1431927613005618.

[24] D.J. Prior, Problems in determining the misorientation axes, for small angular misorientations, using electron backscatter diffraction in the SEM, J. Microsc. 195 (1999) 217–225. https://doi.org/10.1046/j.1365-2818.1999.00572.x.

[25] A.J. Wilkinson, A new method for determining small misorientations from electron back scatter diffraction patterns, Scr. Mater. 44 (2001) 2379–2385. https://doi.org/10.1016/S1359-6462(01)00943-5.

[26] F.R.N. Nabarro, S.B. Luyckx, U. V. Waghmare, Slip in tungsten monocarbide. I. Some experimental observations, Mater. Sci. Eng. A. 483–484 (2008) 139–142. https://doi.org/10.1016/j.msea.2006.09.153.

[27] F.R.N. Nabarro, S.B. Luyckx, U. V. Waghmare, Slip in tungsten monocarbide. II. A first-principles study, Mater. Sci. Eng. A. 483–484 (2008) 9–12. https://doi.org/10.1016/j.msea.2006.09.174.

[28] N.C. Krieger Lassen, Automatic Determination of Crystal Orientation from Electron Backscattering Patterns, Mater. Sci. Doctor of (1994).

[29] S.-S. Rui, L.-S. Niu, H.-J. Shi, S. Wei, C.C. Tasan, Diffraction-based misorientation mapping: A continuum mechanics description, J. Mech. Phys. Solids. 133 (2019) 103709. https://doi.org/10.1016/j.jmps.2019.103709.

[30] W.F. Hosford, On Orientation Changes Accompanying Slip and Twinning, Texture Cryst Solids. 2 (1977) 175–182. https://doi.org/10.1155/TSM.2.175.

[31] P. Chen, S.C. Mao, Y. Liu, F. Wang, Y.F. Zhang, Z. Zhang, X.D. Han, In-situ EBSD study of the active slip systems and lattice rotation behavior of surface grains in aluminum alloy during tensile deformation, Mater. Sci. Eng. A. 580 (2013) 114–124. https://doi.org/10.1016/j.msea.2013.05.046.

[32] K. Thomsen, K. Mehnert, P.W. Trimby, A. Gholinia, Quaternion-based disorientation coloring of

[33] G. Nolze, R. Hielscher, Orientations – perfectly colored, J. Appl. Crystallogr. 49 (2016) 1786–1802. https://doi.org/10.1107/s1600576716012942.

[34] H.G. Jones, Uncertainties in focused ion beam characterisation, University of Surrey, 2020. https://doi.org/10.15126/thesis.00853346.

[35] D. Raabe, D. Ma, F. Roters, Effects of initial orientation, sample geometry and friction on anisotropy and crystallographic orientation changes in single crystal microcompression deformation: A crystal plasticity finite element study, Acta Mater. 55 (2007) 4567–4583. https://doi.org/10.1016/j.actamat.2007.04.023.

[36] T. Csanádi, M. Bl'Anda, N.Q. Chinh, P. Hvizdoš, J. Dusza, Orientation-dependent hardness and nanoindentation-induced deformation mechanisms of WC crystals, Acta Mater. 83 (2015) 397–407. https://doi.org/10.1016/j.actamat.2014.09.048.

[37] G. Nolze, Image distortions in SEM and their influences on EBSD measurements, Ultramicroscopy. 107 (2007) 172–183. https://doi.org/10.1016/j.ultramic.2006.07.003.

[38] R. Hielscher, MTEX 5.2.8, (2020). http://mtex-toolbox.github.io/ (accessed May 25, 2018).

[39] G. Nolze, C. Grosse, A. Winkelmann, Kikuchi pattern analysis of noncentrosymmetric crystals, J. Appl. Crystallogr. 48 (2015) 1405–1419. https://doi.org/10.1107/S1600576715014016.

[40] G. Taylor, M. Farren, The distortion of crystals of aluminium under compression.—Part I, Proc. R. Soc. London. Ser. A, Contain. Pap. a Math. Phys. Character. 111 (1926) 529–551. https://doi.org/10.1098/rspa.1926.0080.

[41] G. Taylor, The distortion of crystals of aluminium under compression. Part II.—Distortion by double slipping and changes in orientation of crystals axes during compression, Proc. R. Soc. London. Ser. A, Contain. Pap. a Math. Phys. Character. 116 (1927) 16–38. https://doi.org/10.1098/rspa.1927.0120.

[42] E.W. Weisstein, Sphere Point Picking., Wolfram MathWorld. (2020). https://mathworld.wolfram.com/SpherePointPicking.html (accessed August 27, 2020).

[43] A. Winkelmann, B.M. Jablon, V.S. Tong, C. Trager-Cowan, K.P. Mingard, Improving EBSD precision by orientation refinement with full pattern matching, J. Microsc. 277 (2020) 79–92. https://doi.org/10.1111/jmi.12870.

# Supplementary Information

# 1 EBSD analysis methods

## 1.1 EBSD data postprocessing

EBSD maps were exported into MTEX [38] for all postprocessing and analysis steps. Each map was rotated 180° in the sample plane so the top of the pillar was at the top of the map, and the compression direction to point approximately along +Y (down in SEM image), since the pre-tilted holder used for EBSD did not allow in-plane sample rotations using SEM stage movement.

The pillar long axes appear slightly rotated away from the EBSD map Y-axes, and is most prominent in the 600 °C prismatic pillar (Figure 4a). This misalignment was not corrected during postprocessing of the EBSD maps, because it comes from a combination of both in-plane and out-of-plane sample rotations, and therefore it would be incorrect to apply a single in-plane rotation correction to realign the pillar with the EBSD map axes. Estimated sample misalignments and EBSD angular resolutions are calculated in Section 7.1.

The space group of hexagonal WC is P-6m2, which does not have inversion symmetry, i.e. $[uvtw]$ is not symmetrically equivalent to $[\bar{u}\bar{v}\bar{t}\bar{w}]$. However, conventional EBSD indexing is insensitive to inversion asymmetry [39], therefore the data are plotted with the symmetry of the equivalent centrosymmetric (Laue) group P6/mmm.

## 1.2 GROD calculation and plotting

EBSD maps were manually segmented into undeformed and deformed regions. The deformed region included the pillar, as well as parts of the base directly under the pillar. The undeformed crystal orientations were measured from map regions next to the pillars.

GROD calculations were performed with respect to the mean orientation of the undeformed map regions. Mean orientations were computed in MTEX [38], which uses the 1$^{st}$ order spherical harmonic coefficient of an orientation cluster's orientation density function.

GROD can be expressed as a rotation with angle and axis components. The rotation angle (Figure 5) was straightforward to plot as it is a scalar. However, the rotation axis more complicated to plot: it is a 3-dimensional unit vector, and can be expressed either in crystal (Figure 6) or in sample coordinates (Figure 7). In Section 0, GROD axes (maps in Figure 5 to Figure 7) were plotted using a colour key which was based on inverse pole figure (IPF) directions, and also accounted for degeneracy in GROD axes at small angles [32]. An explanation of each colour key has been included alongside the results in Section 0.

Pole figures of the GROD rotation axes were also plotted alongside the GROD maps in Figure 6 and Figure 7. Unlike the GROD maps, pole figures do not contain spatially resolved information, but can quantitiatively plot directions, unlike EBSD IPF map colours. Pole density functions were calculated from GROD rotation axes within the pillars using a 2° kernel half-width.

## 1.3 Determining the primary slip system from lattice rotation axes

One of the main aims of this study is to distinguish between $\langle c + a \rangle$ versus $\langle a \rangle$ or $\langle c \rangle$ dislocation slip activity. The following sections describe first the general theory of how the active primary slip systems are determined from GROD rotation axes of deformed micropillars, then how it is applied to slip systems in WC, including special cases of multiple slip.

### 1.3.1 Theory

Plastic deformation is accommodated by shear on discrete crystallographic slip systems of dislocations moving along the slip direction $\boldsymbol{b}$ and in the slip plane $\hat{\boldsymbol{n}}$.

The rotation of a micropillar sample in compression can be described by Taylor's model of single crystal compression. In single slip, the compression plane rotates towards the slip plane, and the axis of rotation is the intersection between the slip plane and compression plane [40,41]. This model has been used in the present study to interpret the GROD axes in sample coordinates. The characteristic rotation axis can also be expressed in terms of the active dislocation slip system $\boldsymbol{\omega} = (\hat{\boldsymbol{n}} \times \boldsymbol{b})$, which is independent of the compression plane or any other sample reference coordinates [30]. However, geometrically unconstrained slip, such as simple shear deformation of a single crystal, or deformation of surface grains in a polycrystal, rotates the deforming object, but does not change the crystal lattice [31].

During micropillar compression, the micropillar is principally constrained in the compression plane. The pillar top face is geometrically fully constrained by the flat indenter tip in the vertical direction, and partially constrained by tip friction in the horizontal direction. The pillar base is horizontally and vertically constrained by chemical bonding to the base material, and the pillar side walls are not constrained. This model will be illustrated for one pillar in Section 4.3 and Figure 8.

For multiple slip, the total lattice rotation is the product of rotations from all locally activated slip systems. In general, the rotation axis cannot be decomposed into its constituent slip systems, since the rotation is path-dependent and depends on which slip systems activated first. If there is one dominant slip system, rotation due to minor activation of secondary slip systems can be treated as 'noise' in the measured rotation axis.

### 1.3.2 Application to WC dislocation slip

The characteristic rotation axes for $\langle a \rangle$, $\langle c \rangle$ and $\langle c + a \rangle$ prismatic single slip systems in WC are $[0001]$ (IPF red), $\langle 11\bar{2}0 \rangle$ (IPF green), and $\langle 1\ 1\ \bar{2}\ 3.15 \rangle$ (IPF orange) respectively.

Multiple $\langle a \rangle$ prismatic slip is a special case where the lattice rotation axis is shared between all symmetric variants, so both single and multiple $\langle a \rangle$ prismatic slip have a rotation axis of $[0001]$. If slip on two $\langle a \rangle$ prismatic systems causes opposing lattice rotations, they will cancel out and the GROD angle will be small, but the rotation axis will still be parallel to $[0001]$.

Cooperative multiple $\langle c + a \rangle$ slip on a single prismatic plane will move the GROD axis between $[0001]$ and $\langle 11\bar{2}0 \rangle$ along the red-orange-green line in Figure 3 depending on applied load, as the total slip direction can lie anywhere in the prismatic plane. However, if multiple prismatic slip planes are activated, the total slip direction will not lie in a single prismatic plane, so the total GROD axis will deviate from the red-orange-green line.

# 2 Misorientation axis uncertainty

## 2.1 Theoretical misorientation axis error calculation

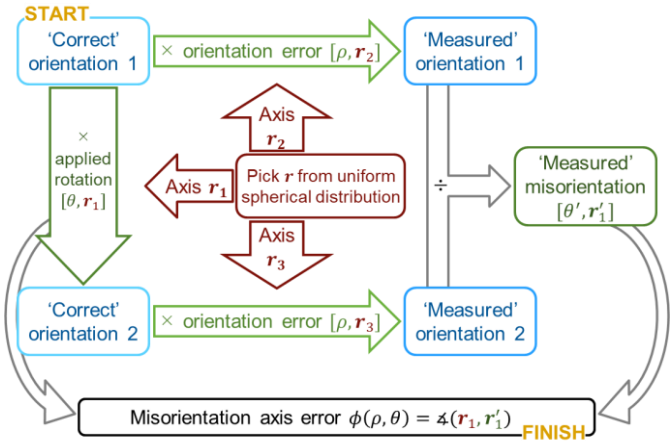

*Supplementary Figure 1: Flowchart showing how misorientation axis error ϕ is calculated for a given orientation error ρ and applied rotation angle θ. Rotations are described as [angle, axis] pairs.*

Each box and whisker plot in Figure 11e showed the theoretical misorientation axis uncertainty distribution $\phi$ for a given orientation error $\rho$ and rotation angle $\theta$. The flowchart in Supplementary Figure 1 shows the method for how this was calculated.

First, a known rotation $(\theta, r_1)$ is applied to an arbitrary starting orientation ('Correct orientation 1') to obtain a second orientation ('Correct orientation 2'). Then, simulated measurement errors, $(\rho, r_2)$ and $(\rho, r_3)$, are applied to the two orientations to obtain 'Measured orientation 1' and 'Measured orientation 2' respectively. The misorientation between 'Measured orientation 1' and 'Measured orientation 2' is $(\theta', r_1')$. Any deviation between the measured misorientation $(\theta', r_1')$ and applied rotation $(\theta, r_1)$ is due to the orientations 'wobbling' from measurement error. The misorientation axis error $\phi$ is the angle between the applied misorientation axis $r_1$, and the measured axis $r_1'$.

The exact value of $\phi$ depends on the measurement error axes $r_2$ and $r_3$, so the calculation is repeated 400 times for each $\rho$ and $\theta$ to obtain distributions of $\phi(\rho, \theta)$ in Figure 11e. In all cases, each rotation axis $r_i$ is a random unit vector picked from a uniform spherical distribution [42]. This is probably a reasonable approximation for EBSD patterns acquired with a large capture angle and uniform background intensities.

Activation of secondary slip systems will also change the lattice rotation axis. Quantitative assessment of secondary slip contributions is generally not possible, since the deviation is path dependent. However, secondary slip contributions are expected to systematically bias the orientation error more strongly than orientation indexing errors from band localisation or EBSD pattern brightness.

The misorientation axis errors reported in Reference [25] are overlaid onto the theoretical plots in Figure 11e, where average and maximum lines nominally correspond to orientation uncertainties of 0.5° and 1.4° respectively. The misorientation axis errors in Reference [25] agree with the theoretical distributions.

## 2.2 Choice of EBSD orientation uncertainty metric

Three metrics for EBSD angular resolution were considered: MAD, KAM and spatially uncorrelated misorientations. Figure 11a-d shows histogram distributions of all three EBSD uncertainty metrics.

MAD measures the average deviation between detected and indexed band positions in an EBSD pattern. MAD is an upper bound estimate of uncertainty, because the solved orientation, which has three degrees of freedom, is the best fit from typically 10-12 Kikuchi bands (7 bands minimum).

KAM is a lower-bound estimate of EBSD orientation uncertainty, because any bias in orientation error will be cancelled out in the misorientation calculation. It also presumes that there are no lattice rotations in the sample, which is fine for undeformed tungsten carbide, since short-range lattice rotations are negligible relative to EBSD orientation precision using the Refined Accuracy method [43]. The four nearest neighbours of each map point were used for KAM calculation.

The spatially uncorrelated misorientation distribution lies somewhere in between KAM and MAD. Spatially localised indexing bias, e.g. from sample surface topography, are not cancelled out (as they can be in KAM), but other sources of indexing bias, such as uncertainty in EBSD pattern centre, could still cancel out. The spatially uncorrelated misorientation distributions were calculated from at least 176000 pairs of randomly selected EBSD map points (ten times the number of datapoints). For this data, the uncorrelated misorientation distributions were used to measure the orientation uncertainty. The three prismatic oriented pillars show a median orientation scatter of about 0.2° and 95$^{th}$ percentile of 0.5°.

The misorientation distribution for the basal pillar was higher than the other three pillars (0.27° average / 0.9° 95$^{th}$ percentile), possibly due to the noisier regions on the right edge of the mapped region far from the WC pillar, with higher misorientation angles most likely caused by FIB-sectioning topography artefacts (Figure 11e inset). Fortunately, the lower angular resolution for this pillar is less critical for misorientation analysis due to its large lattice rotations of 5-30° (Figure 5b).

# 3   Additional Supplementary Figures

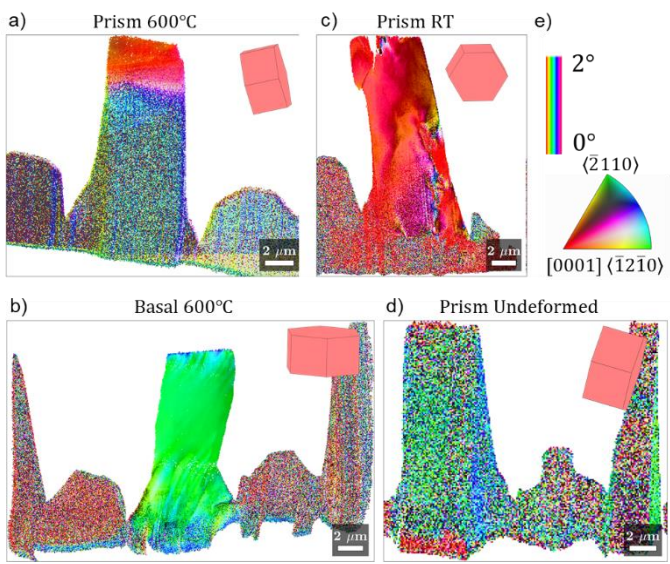

*Supplementary Figure 2: GROD axis only, crystal coordinates (no misorientation angle)*

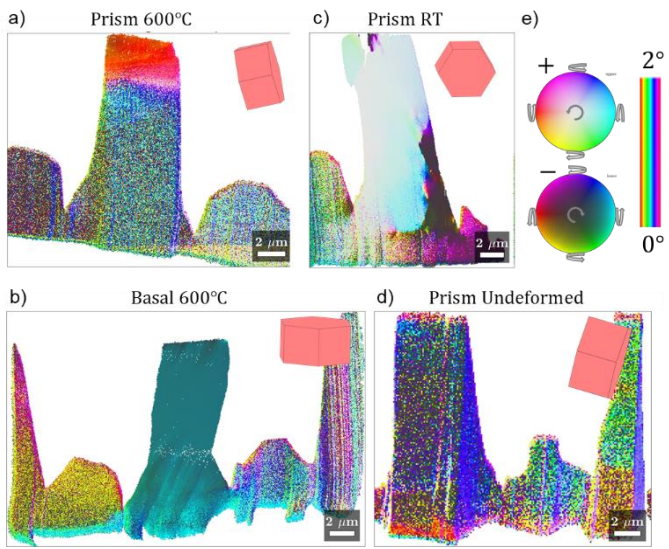

*Supplementary Figure 3: GROD axis only, sample coordinates (no misorientation angle)*